\documentclass{osa-article}

\journal{osajournal}

\articletype{Research Article}

\usepackage{graphicx}
\usepackage{dcolumn}
\usepackage{bm}
\usepackage{xcolor}
\DeclareMathOperator{\argminG}{arg\,min} 
\DeclareMathOperator{\argmaxG}{arg\,max} 

\begin{document}

\title{Photonic extreme learning machine by free-space optical propagation}

\author{Davide Pierangeli\authormark{1,2}, Giulia Marcucci\authormark{3}, and Claudio Conti\authormark{2,1}}
\address{\authormark{1}Institute for Complex System, National Research Council (ISC-CNR), 00185 Rome, Italy\\
\authormark{2} Physics Department, Sapienza University of Rome, 00185 Rome, Italy\\
\authormark{3} Department of Physics, University of Ottawa, Ottawa ON K1N 6N5, Canada\\}
\email{\authormark{*}davide.pierangeli@roma1.infn.it}

\begin{abstract}
Photonic brain-inspired platforms are emerging as novel analog computing devices, enabling fast and energy-efficient operations 
for machine learning. These artificial neural networks generally require tailored optical elements, such as 
integrated photonic circuits, engineered diffractive layers, nanophotonic materials, or time-delay schemes, 
which are challenging to train or stabilize.
Here we present a neuromorphic photonic scheme - photonic extreme learning machines - that can be implemented simply by using an optical encoder
and coherent wave propagation in free space. We realize the concept through spatial light modulation of a laser beam, with the far field 
that acts as feature mapping space. We experimentally demonstrated learning from data on various classification and regression tasks, 
achieving accuracies comparable to digital extreme learning machines.
Our findings point out an optical machine learning device that is easy-to-train, energetically efficient, scalable and fabrication-constraint free.
The scheme can be generalized to a plethora of photonic systems,
opening the route to real-time neuromorphic processing of optical data.
\end{abstract}

\section{INTRODUCTION}

Artificial neural networks excel in learning from data to perform classification, regression and prediction tasks of vital importance
\cite{Haykin2008}. As data information content increases, simulating these models becomes computationally expensive on conventional computers,
making specialized signal processors crucial for intelligent systems.
Training large networks is also very costly in terms of energy consumption and carbon footprint \cite{Strubell2019}.
Photonics provides a valuable alternative towards sustainable computing technologies. For this reason,  
machine learning through photonic components is gathering an enormous interest \cite{Wetzstein2020}.
In fact, many mathematical functions, which enables complex features extraction from data, find a native implementation on optical platforms.
Pioneering attempts using photosensitive masks \cite{Farhat1985} and volume holograms \cite{Psaltis1988, Denz1998} have been developed very recently
into coherent optical neural networks that promise fast and energy-efficient optical computing \cite{Shen2017, Lin2018, Engheta2019, Bueno2018, Zuo2019, Tait2017, Feldmann2019, Feldmann2021, Xu2021, Moss2020}.
These schemes exploits optical units such as nanophotonic circuits \cite{Shen2017}, on-chip frequency combs \cite{Feldmann2021, Xu2021, Moss2020}, 
and spatial light modulators to perform matrix multiplications in parallel \cite{Lvovsky2020}, or to carry out convolutional layers \cite{Chang2018, Miscuglio2020, Wu2021}. Training consists in adjusting the optical response of each physical node \cite{Hughes2019}, also adopting external optical signals \cite{Hamerly2019}, and it is very demanding \cite{Hughes2018}. 
Moreover, photonic neural networks based on nano-fabrication still have a considerable energy impact.
A general and promising idea to overcome the issue is adapting machine-learning paradigms and make them inclined to optical platforms.
In this article, we pursue this approach by constructing an easy-to-train optical architecture that require only free-space optical propagation.

Photonic architectures that do not need control of the entire network are particularly attractive.
A remarkable method for their design and training is reservoir computing \cite{Lukosevicius2009, Gallicchio2017, Mattheakis2017}, 
in which the nonlinear dynamics of a recurrent system processes data, and training occurs only on the readout layer.
Optical reservoir computing has demonstrated striking performance on dynamical series prediction by using delay-based setups \cite{Brunner2013, Vinkier2015, Larger2017, Antonik2019}, laser networks \cite{Rohm2020}, multimode waveguides \cite{Paudel2020}, and random media \cite{Rafayelyan2020}.
Since several interesting complex systems can be exploited as physical reservoirs, the inverse approach is also appealing: 
the scheme can be trained for learning dynamical properties of the reservoir itself \cite{Pierangeli2020}.

Extreme learning machines (ELMs) \cite{Huang2006, Huang2012}, or closely related schemes based on random neural networks~\cite{Schmidt1992, Pao1994}, 
support vector machines \cite{Suykens1999}, and kernel methods \cite{An2007}, are a powerful paradigm in which only a subset of connections is trained.
The basic mechanism is using the network to establish a nonlinear mapping between the dataset and a high-dimensional feature space,
where a properly trained classifier performs the separation. 
In optics, an interesting case of this general approach has been implemented 
by using speckle patterns emerging from multiple scattering ~\cite{Saade2016} and multimode fibers~\cite{Uchida2020} as a feature mapping. 
Although in principle many optical phenomena can form the building block of the architecture~\cite{Marcucci2020, Psaltis2020}, 
the general potentials of the ELM framework for photonic computing remains largely unexplored.

Here, we propose a photonic extreme learning machine (PELM) for performing classification and regression tasks using coherent light.
We find that a minimal optical setting composed by an input element, which encodes embedded input data into the optical field, a
wave layer, corresponding to propagation in free space, and a nonlinear detector, enables simple and efficient learning.
The encoding method has a crucial role in determining the learning ability of the machine.
The architecture is experimentally implemented via phase modulation of a visible laser beam by a spatial light modulator (SLM).
We benchmark the realized device on problems of different classes, achieving performance comparable with digital ELMs.
These includes a classification accuracy exceeding 92\% on the well-known MNIST database, which overcomes
diffractive neural networks fabricated after an onerous training \cite{Lin2018}.
Given the massive parallelism provided by spatial optics and the ease of training, 
our approach is ideal for big data, i.e., extensive datasets with large dimension samples. 
Our PELM is intrinsically stable and adaptable over time, as it does not require engineered or sensitive components. 
It can potentially operate in dynamic environments as intelligent device performing on-the-fly optical signal processing.

\begin{figure*}[t!]
\centering
\vspace*{-0.1cm}
\hspace*{-0.4cm} 
\includegraphics[width=0.7\columnwidth]{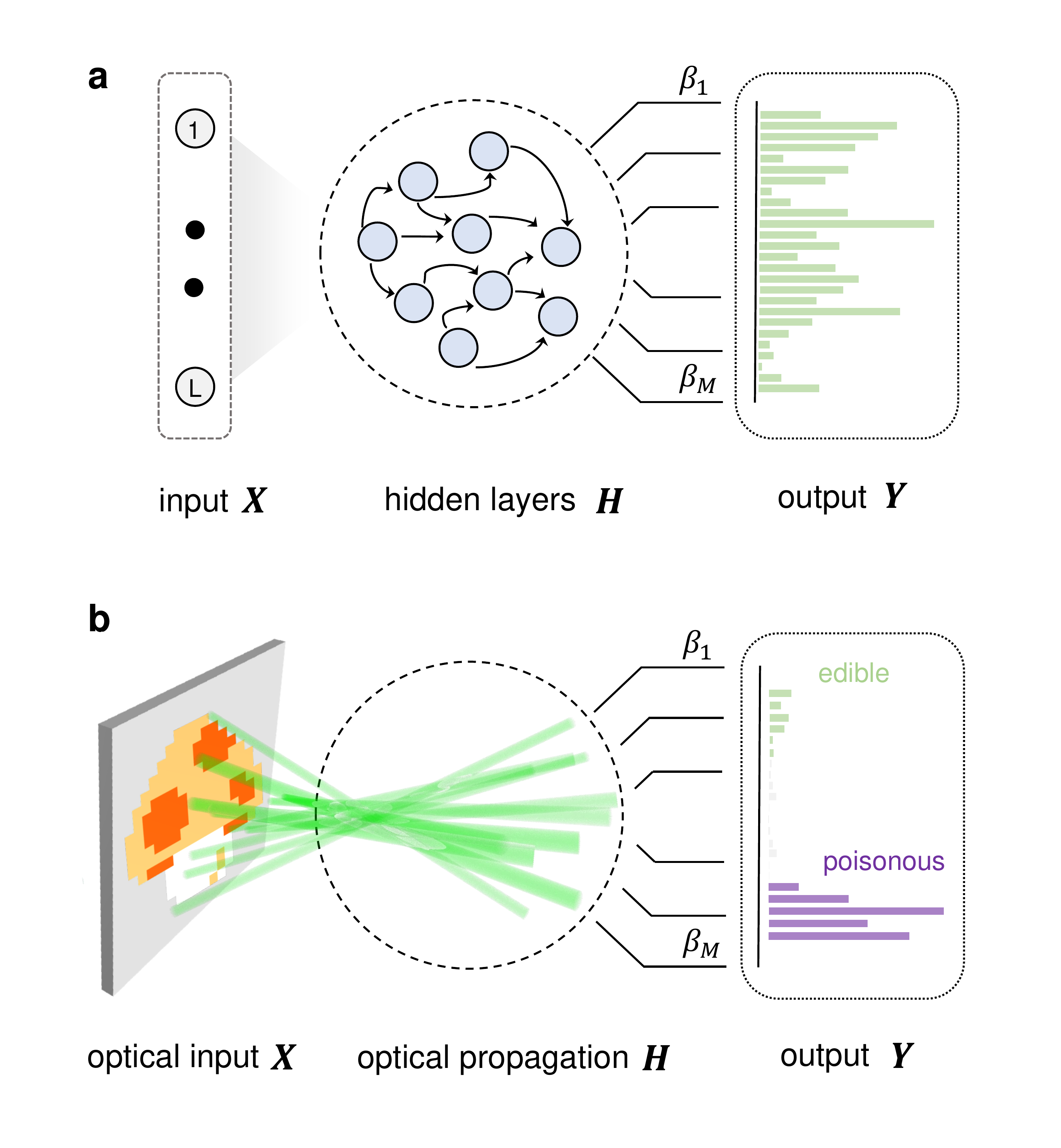} 
\vspace*{-0.3cm}
\caption{{Schematic architecture of the photonic extreme learning machine (PELM).} 
(a) A general ELM scheme with the input dataset $\mathbf{X}$ that is fed into a reservoir, which 
gives out the hidden-layer output matrix $\mathbf{H}$. The trainable readout weights $\mathbf{\beta}_i$ determine the network output $\mathbf{Y}$.
(b) In the optical case, the input (a mushroom in the example) is encoded on the optical field and hidden neurons have been replaced by modes that interact during propagation. Training of the photonic classifier is enabled by $M$ detection channels. 
}
\vspace{-0.1cm}
\label{Figure1}
\end{figure*}

\section{PELM ARCHITECTURE}

An ELM is a feed-forward neural network in which a set of input signals are connected to at least one output node by
a large structure of generalized artificial neurons~\cite{Huang2006}. An outline of the architecture is illustrated in Fig.~1(a). 
The hidden neurons form the computing reservoir. Unlike from neurons in a deep neural network, they do not require any training and their
individual responses can be unknown to the user~\cite{Huang2012}. 
Given an input dataset with $N$ samples $\mathbf{X}=[\mathbf{X}_1, ..., \mathbf{X}_N]$, the reservoir furnishes an hidden-layer
output matrix $\mathbf{H}=[\mathbf{g}(\mathbf{X}_1), ..., \mathbf{g}(\mathbf{X}_N) ]$, where $\mathbf{g}(\mathbf{x})$ is a nonlinear feature mapping. 
$\mathbf{H}$ is linearly combined with a set of weights \textbf{$\beta$} to give the output $\mathbf{Y}$ performing the classification,
$\mathbf{Y}=\mathbf{H}\beta$.
To train an ELM classifier, the optimal output weights $\beta_i$ are the sole parameters that need to be determined, a problem that can be solved via ridge regression. Details on this technique are reported in Appendix A.

We transfer the ELM principle into the optical domain by considering the three-layer structure illustrated in Fig.~1(b).
In the encoding layer, the input vectors $\mathbf{X}_i$ are embedded into the phase and/or amplitude of the field by an optical modulator.
The reservoir consists of linear optical propagation and nonlinear detection of the final state.
The output is recovered in the readout layer, where weights $\beta_i$ are applied to $M$ measured channels.
The $\beta$ set is trained by solving the regression problem via digital hardware.
For an extensive training set, with size larger than the number of channels ($N \gg M$), an effective solution reads as~\cite{Huang2012}
\begin{equation}
\beta= \left( \mathbf{H^{T}}\mathbf{H} +  c \mathbf{I} \right)^{-1} \mathbf{H^{T}} \mathbf{T}
  \label{eq:1}
\end{equation}
where $\mathbf{T}$ indicates the training targets, $\mathbf{I}$ the identity matrix, and $c$ is a regularizing constant.
For a generalized PELM based on the scheme in Fig.~1(b), we construct the feature mapping
\begin{equation}
  {H}_{ji}= \mathbf{g}_i(\mathbf{X}_j)= \mathit{G}  \left( \left[ \mathbf{M} \cdot p( \mathbf{X}_j) \cdot q(\mathbf{W}) \right]_i \right)
  \label{eq:2}
\end{equation}
where $\mathit{G}$ is a detection function, $\mathbf{M}$ is the complex \textit{transfer} matrix modelling field evolution, 
$p$ and $q$ are two \textit{encoding} functions, and  $\mathbf{W}$ is a fixed character of the encoder that we term \textit{embedding} matrix.
$\mathbf{W}$ has no a direct equivalent in the ELM model but emerges in the optical case, where the signal is a complex-valued field.
In such a network, each operation is defined by the corresponding optical elements.
In general, Eq.~(2) can represent the feature space of different optical settings. 
For instance, in the optical classifier demonstrated by \textit{Saade et al.} \cite{Saade2016}, in which amplitude modulated data propagates through a scattering medium, $p(\mathbf{X}_j)=\mathbf{X}_j$ (amplitude encoding) and $\mathbf{M}$ is a random complex Gaussian matrix.
In the three-component system we validate and realize in experiments,  
which is composed by a phase-only SLM, free space, and a camera, Eq.~(2) can be specified as follows. 
Phase encoding of the input data by spatial light modulation corresponds to $p(\mathbf{x})=\exp(i\mathbf{x})$, 
and, since a single encoder is employed, $p=q$, and $\mathbf{H}=\mathit{G} \left( \mathbf{M}\cdot \exp{i( \mathbf{X} + \mathbf{W})}  \right) $.
The nonlinear function $G$ models the detection of the transmitted field.
Using the saturation effect of the camera pixels, we have $G(I)\simeq I/(I+Is)$, with $I=|A|^2$, optical amplitude $A$ and saturation intensity $I_s$.
For free-space optical propagation, i.e., the light distribution in the far field or in the focal plane of a lens, 
$\mathbf{M}$ corresponds to the discrete Fourier transform~\cite{Goodman2005}.

\begin{figure*}[t!]
\centering
\vspace*{-0.4cm}
\hspace*{-0.5cm} 
\includegraphics[width=1.1\columnwidth]{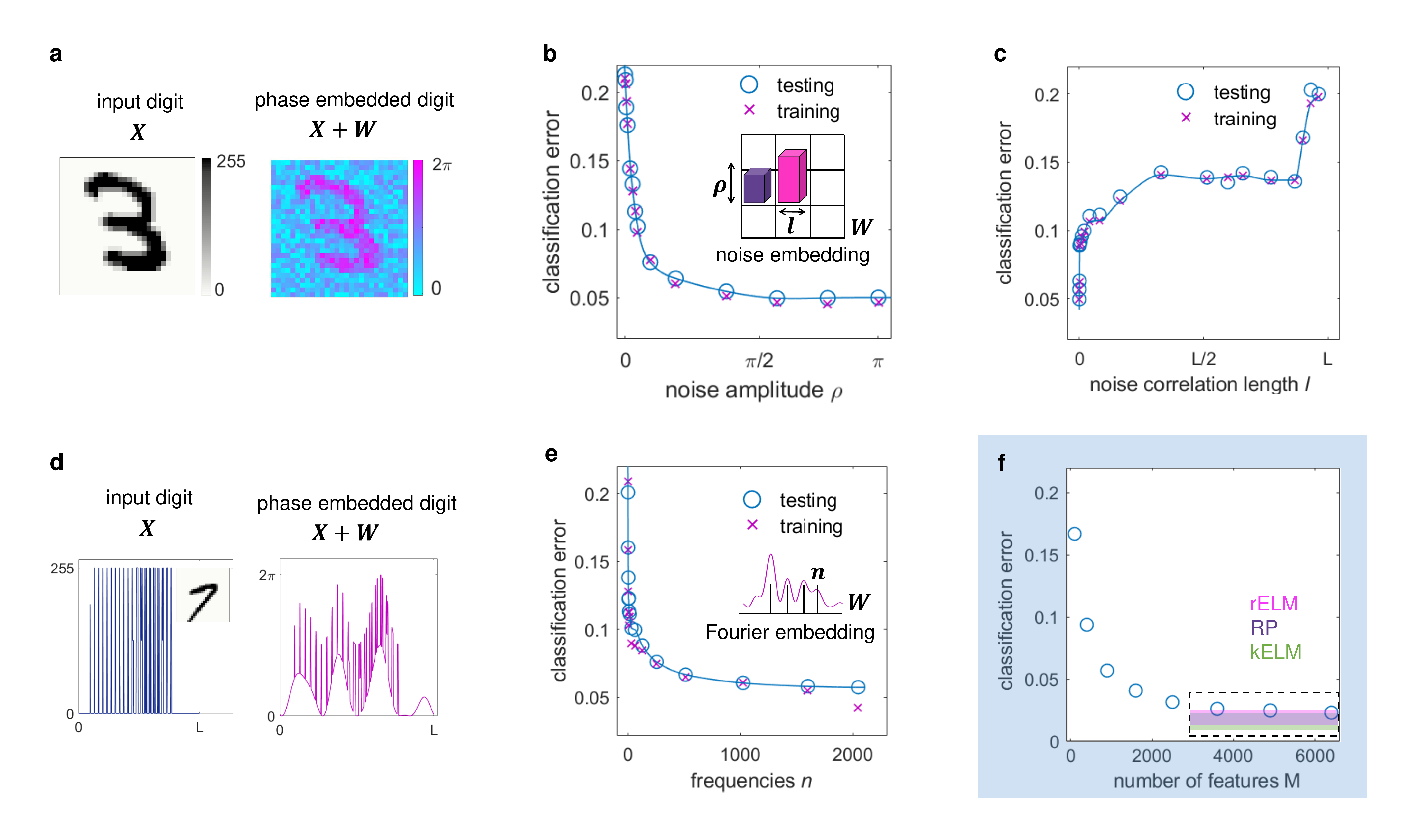} 
\vspace*{-0.3cm}
\caption{ {Learning ability of the PELM architecture.} The optical computing scheme is evaluated on the MNIST dataset by varying the encoding properties and feature space. (a) Input digit and two-dimensional phase mask showing its encoding by noise embedding: the input signal overlaps with a disordered matrix.
(b-c) PELM training and testing error for noise embedding when varying the (b) noise amplitude and (c) its correlation length, for $M=1600$.
(d) Input vector encoded over a carrier signal (Fourier embedding). (e) Classification error versus the number of frequencies of the embedding signal.
(f)~Minimum testing error increasing the number of features $M$. Indicated are the best accuracies reported with random ELM (rELM) \cite{Huang2013}, random projections (RP) \cite{Saade2016} and kernel ELM (kELM) \cite{Huang2013} on the same task.
}
\vspace{-0.1cm}
\label{Figure2}
\end{figure*}

We first validate the free-space PELM architecture and assess the condition for learning via numerical simulations.
We consider digit recognition and we train our classifier on the MNIST handwritten digit database.
Figure 2 reports the learning properties for two representative phase encoding methods: (i) noise embedding and (ii) Fourier embedding.
In (i) $\mathbf{W}$ is a disordered matrix modelling a distortion on the encoder (see Appendix A for details). 
It remains unchanged during both training and testing. 
Each input is encoded by phase modulation in $[0,\pi]$, and the embedding signal is encoded in the same phase interval.
The effect of the noise embedding is illustrated in Fig. 2(a) where a typical digit is shown as a phase mask. 
Figure 2(b) reports the classification performance of the PELM with $M=1600$ channels when varying the mean noise amplitude. 
The machine always reaches accuracies close to that allowed by training.
The classification error shows a sharp decrease as noise increases from zero, and it converges to a plateau already for small perturbations.
Results varying the noise correlation length at fixed noise level are in Fig.~2(c). 
This behavior indicates that optimal learning occurs for small-scale noise, i.e., when $\mathbf{W}$ has its maximum information content (rank $L$).
The opposite occurs for a constant $\mathbf{W}$, with the machine operation that reduces to a straight optical convolution of the input \cite{Miscuglio2020}.
The property is general and extends to embedding methods without randomness. In (ii) $\mathbf{W}$ is a generic carrier signal [Fig. 2(d)].
As shown in Fig. 2(e), a few frequencies in the carrier are sufficient for the learning transition. 
The PELM is much more accurate in the classification as larger is frequency content of the embedding signal.
These results reveal that the embedding matrix has a key role in enabling dataset learning. 
Although in our experiments we control this matrix via the encoder,
we note that it can be intrinsic in any optical setup, for example as a distortion of the optical wavefront.

\begin{figure*}[t!]
\centering
\vspace*{-0.4cm}
\hspace*{-1.0cm}
\includegraphics[width=1.15\columnwidth]{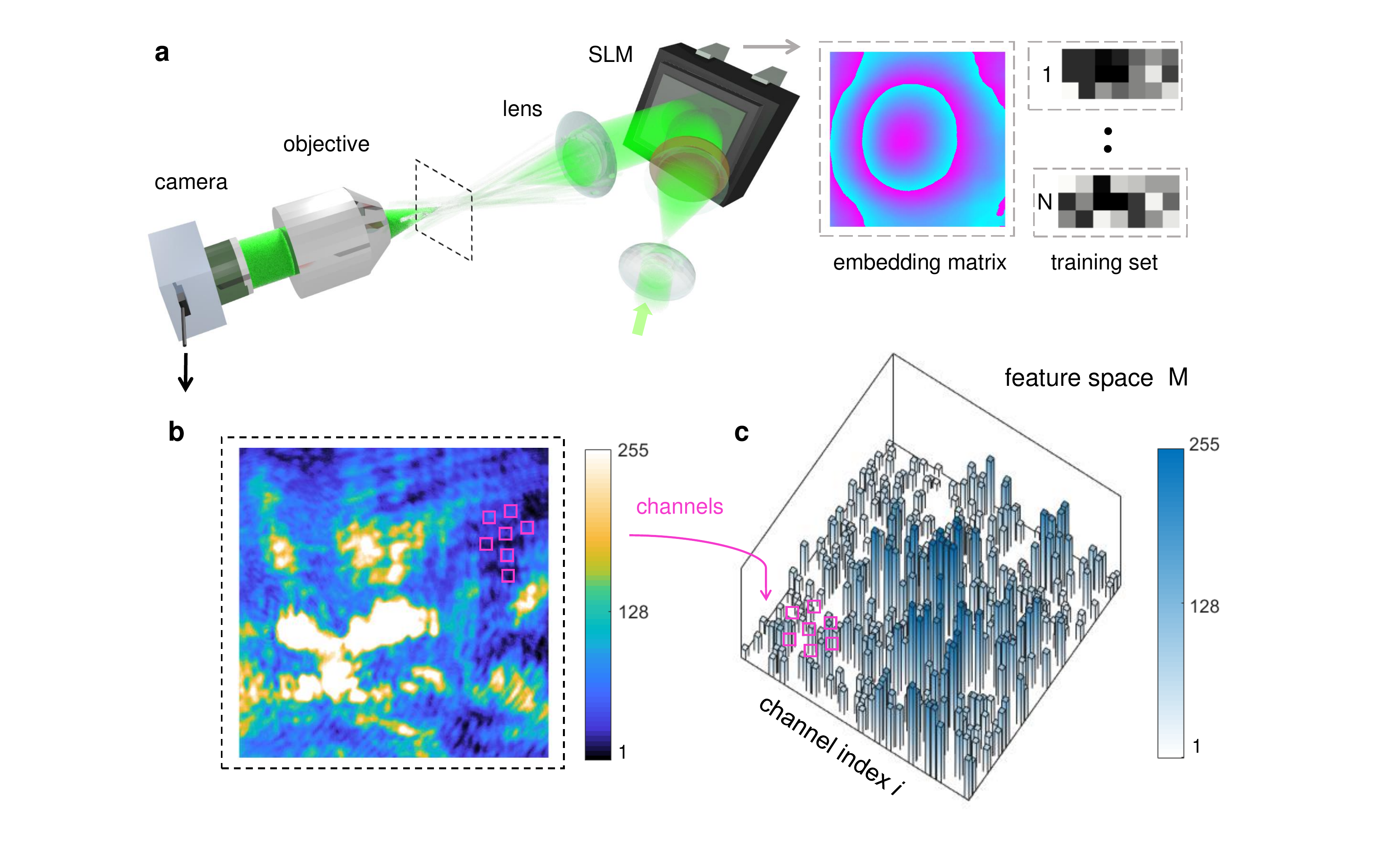} 
\vspace*{-0.3cm}
\caption{{Experimental implementation.} (a) Sketch of the optical setup. 
A phase-only spatial light modulator (SLM) encodes data on the wavefront of a $532$ nm continuous-wave laser. 
The far field in the lens focal plane is imaged on a camera. Insets show a false-colour embedding matrix and training data encoded as phase blocks, respectively. (b) Detected spatial intensity distribution for a given input sample.
White-coloured areas reveal camera saturation in high-intensity regions, which provides the network nonlinear function.
Pink boxes shows some of the $M$ spatial modes (blocks of pixels) that are used as readout channels.
(c) Example of an input data in a feature space of dimension $M=256$, as projected by the optical device.
Each bar represents an output channel, and training consists in finding the vector that properly tunes all the bar heights.
}
\vspace{-0.1cm}
\label{Figure3}
\end{figure*}

\begin{figure*}[t!]
\centering
\vspace*{-0.4cm}
\hspace*{-0.6cm}
\includegraphics[width=1.1\columnwidth]{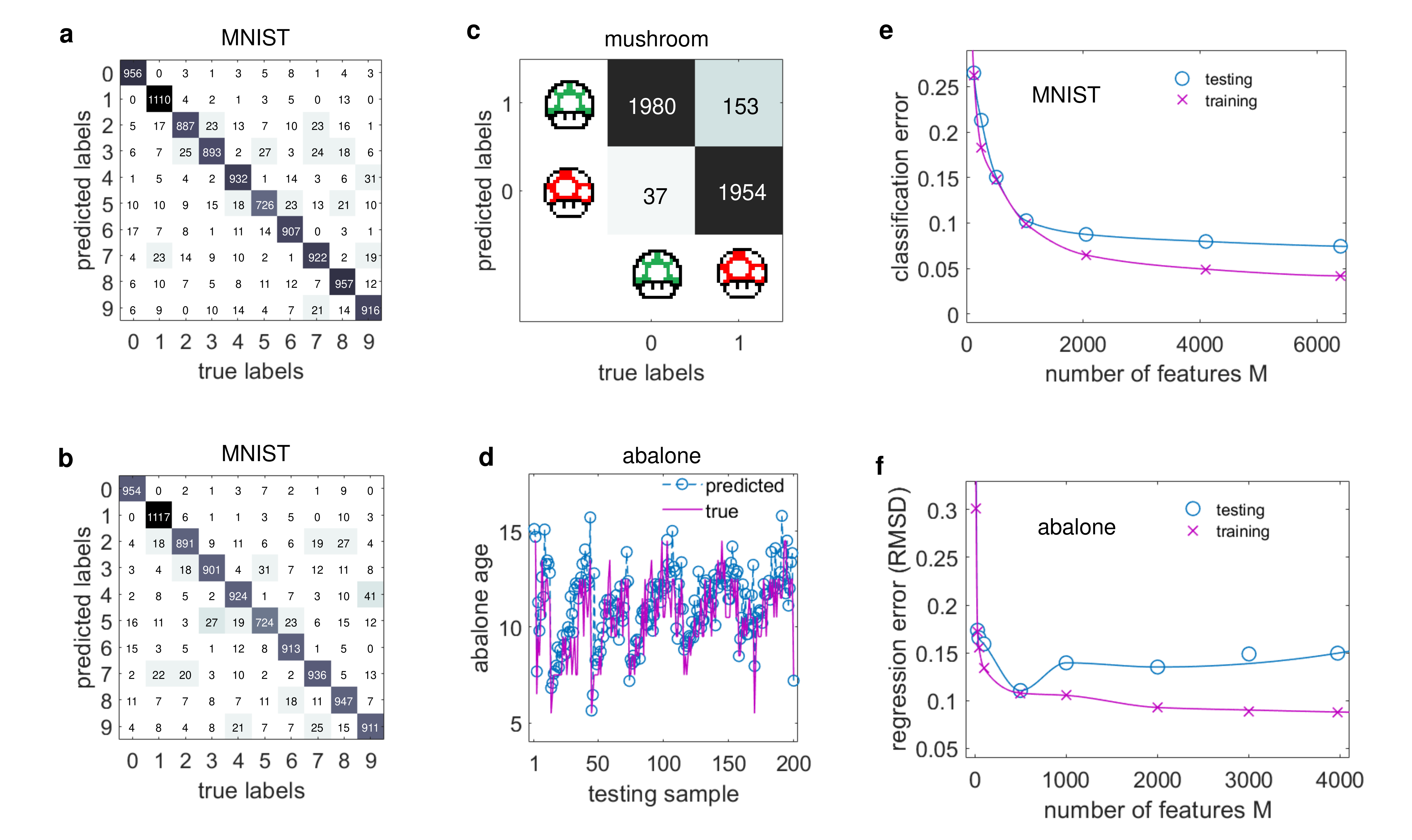} 
\vspace*{-0.2cm}
\caption{ {Experimental performance of the PELM on classification and regression tasks.} Confusion matrices on the MNIST dataset for 
a free-space PELM that makes use of (a) Fourier and (b) random embedding ($92.18$\% and $92.06$\% accuracy, $M=4096$). 
(c) Performance on the mushroom binary classification problem ($95.4$\%). 
(d) Optical predictions and true values for the abalone dataset. 
(e) Classification and (f) regression error as a function of the number of features. 
Rapid convergence to optimal performance is found. Experimental results are compared with numerical simulations and training errors.
}
\vspace{-0.1cm}
\label{Figure4}
\end{figure*}

Another key hyperparameter of the scheme is the number of features (channels) $M$, which set the dimensionality of the feature (measurement) space. 
Figure 2(f) reports the testing error as $M$ increases. The performance rapidly grows with the number of features, whereas 
only hundreds of channels are necessary for a proficient PELM. 
For $M\gtrsim N/20$ the error approaches 2\%, a value very close to the optimal accuracy achievable with learning machines based on the same paradigm \cite{Huang2013}. 

In PELM scheme, the main computational cost for achieving the classification is imputable to large matrix multiplications
and their processing with nonlinear functions. On the optical platform we set up, these operations occurs simply by free-space light propagation
and detection.

\section{EXPERIMENTAL DEVICE}

The experimental setup for the free-space photonic extreme learning machine is illustrated in Fig.~3(a) and detailed in Appendix B. 
A phase-only SLM encodes both the input dataset and the embedding matrix on the phase spatial profile of the laser beam.
In practice, distinct attributes of a dataset are arranged in separate blocks, a data sample corresponds to a set of phase values on these blocks,
and a preselected phase mask is superimposed [inset in Fig.~3(a)].
Phase information self-mixes during light propagation, a linear process that corresponds to the matrix multiplication in Eq.~(2).
The far field is detected on a camera that performs the nonlinear activation function. As shown in the example of Fig.~3(b), 
the resulting acquired image is a saturated function of the optical amplitude in the lens focal plane.
From this data matrix we extract the $M$ channels to construct the PELM feature space. 
An example of an input data measured on a feature space is reported in Fig.~3(c). 
In analogy with the numerical implementation, the device is trained by finding the weights $\mathbb{\beta}$ that, when applied the optical output,
allow performing the classification.

We test the optical device on various computing tasks and datasets. The aim is to point out that, when the PELM is fully effective on a given task, 
it can be also easily and rapidly retrained for a different application.
The main results are summarized in Fig.~4 and demonstrate the learning capability of the optical device.
We first use the MNIST dataset to prove classification on a large-scale multiclass problem.
When trained using $M=4096$ output channels, the PELM reaches a mean classification accuracy of $92$\% $(\pm 0.005)$ over $N_t=10000$ 
test images [Figs. 4(a-b)]. We obtain accuracy exceeding recent optical convolutional processors \cite{Xu2021}, 
and comparable with optical deep neural networks \cite{Lin2018}, but without the fabrication of specific optical layers and the heavy training they require.
The best classification performance is not very sensitive to the embedding matrix employed on the encoder,
which experimentally confirms our numerical findings (Fig. 2). We report in Fig. 4(b) a confusion matrix measured using the random embedding method,
which gives $92.18$\% accuracy, to be compared with the $92.06$\% [Fig. 4(a)] when using Fourier embedding. 
By only updating the input training set and the output weights, the PELM can be quickly reconfigured for classifying different objects. 
We consider a binary class problem, the mushroom task, where the goal is to separate poisonous from edible mushrooms 
from a series of features (see Appendix C for details).  Figure 4(c) shows that on this dataset we achieve $95.4$\% accuracy, using $N_t=4124$ test samples
and $N=4000$ training points, which is very close to the precision of digital ELMs \cite{Huang2012}.

The excellent properties of the experimental device as a classifier generalizes to regression problems.
We test the abalone dataset (see Appendix C), a task where the input data ciphers a sea snail and the output is expected to furnish its age.
It is a solid benchmark since has low-dimensional inputs ($L=8$) but contains extreme events that are difficult to predict. 
Figure 4(d) shows experimental abalone predictions obtained via a feature space of dimension $M=2000$.
The testing error measured as normalized root mean square displacement (NRMSD) is $\approx 0.12$, i.e., regression is performed with remarkable accuracy. 

A key point of the PELM architecture is a testing error that rapidly converges to its optimal value as the feature space dimension increases.
Experimental results demonstrating this property are in Figs. 4(e-f), respectively for the MNIST and abalone dataset.
Good classification/regression performance is maintained even for a very small number of trained channels, if compared to the dataset size.
For instance, useful abalone predictions can be obtained with only $M=128$ channels, i.e., with a training process that consists only in inverting a modest matrix ($128^2$ elements). Heuristically, we find that the accuracy reaches a limiting values as $M\sim N/20$, in agreement with numerical simulation 
for the same machine hyperparameters. The same behavior is found for training errors, which indicates that input data are already completely separated by the optical setup.

\section{DISCUSSION}

Training errors in Figs. 4(e-f) allow us to identify the various factors towards improving the computational resolution of the optical machine.
We ascribe residual errors to practical non-idealities of the device. 
The finite 8-bit precision of both encoding and readout, and noise in optical intensities, 
introduce errors that are absent in the digital implementation.  
Flickering effects of the phase modulator, which give a tiny but variable inaccuracy each time the machine is interrogated, 
are identified as one of the dominant source of error in the realized classifier.
On the other hand, small optical and mechanical fluctuations of the tabletop optical line 
can result in further inconsistencies during training. These fluctuations can be compensated in future experiments using incremental learning \cite{Huang2012}, a refined technique that allows to adaptively adjust the weights while the training is ongoing.

Besides a computing effectiveness comparable to its digital counterpart, the PELM hardware offers several advantages that are promising
for fast processing of big data, especially for inputs that naturally present themselves as optical signals.
First, unlike optical deep neural networks \cite{Yan2019, Luo2019}, the leading computational cost for training the device does not depend 
on the dataset size $N$, but only on the selected number of channels $M$. 
Once trained, forward information propagates optically in a fully passive way and in free space.
This can provide a key advantage with respect to ELM algorithms in terms of both scalability and latency.
In fact, the matrix multiplication in Eq. (2), which maps input data to the feature space, on a digital computer 
requires a time and memory consumption that grow quadratically with the data dimension $L$. 
The optical hardware performs this operation totally in parallel, independently of the input data size, 
and without power dissipation. 
Since only $M$ scalar multiplications are necessary from the optical detection to the final classification, the PELM has a very low latency,
which is independent on the dataset dimension. This is a crucial property in applications that require fast responses,
such as in real-time computer vision.
The main speed limitation of our free-space PELM is related to the frame rate of the input optical modulator. 
Liquid-crystal SLMs currently have typical frame rates of the order of $100$ Hz, but phase modulators based on micro-electro-mechanical, electro-optical or  acousto-optical technologies are approaching GHz frequencies \cite{Tzang2019, Braverman2020}. 
With similar components available, PELM hardware could perform $R=L \times M \times 10^9$ operation per seconds (OPS),
a value that for large $L$ can overcome the current limits of electronic computing (peta OPS).

Finally, the absence of any optical element or medium along the neural network path, which differentiates our scheme from all other optical neuromorphic devices previously realized, is a valuable aspect for various reasons.  The tiny optical absorption and scattering of coherent light in air implies that our scheme needs low-power lasers ($\mu$W or lower), i.e., it requires minimal energy consumption.
More importantly, our device is extremely stable to mechanical and optical external perturbations, because its operation does not depend on components 
that must be kept completely static after training. This suggests that the proposed optical processor is promising for applications in dynamic environments.

Nowadays artificial intelligence is permeating the field of photonics \cite{Piccinotti2020, Genty2020}, and viceversa \cite{Kutz2017}. 
Optical platforms are enabling computing functionalities with unique features,
ranging from photonic Ising machines for complex optimization problems \cite{McMahon2016, Inagaki2016, Vandersande2019, Pierangeli2019, Pierangeli2020_2, Prabhu2020, Kalinin2020, Gaeta2020} to optical devices for cryptography \cite{DiFalco2019}.
We have realized a novel photonic neuromorphic computing device that is able to perform classification and feature extraction only  
by exploiting optical propagation in free space. 
Our results point out that fabricated optical networks, or complex physical processes and materials, 
are not mandatory ingredients for performing effective machine learning on a optical setup.
All the essential factors for learning can be included in the encoding and decoding method by exploiting natural field propagation.
On this principle, we demonstrated a photonic extreme learning machine that, given its unique adaptability and intrinsic stability,
is attracting for future processing of optical data in real-time and dynamic conditions.
More generally, our approach envisions the exceptional possibility of harnessing any wave system as an intelligent device that learn from data, 
in photonics and far beyond.

\section{APPENDIX A: NETWORK DETAILS}


\subsection{ELM framework}
 
The basic ELM structure is a single-layer feedforward neural networks in which hidden nodes are not tuned	\cite{Huang2006}. 
For real-valued random internal weights, the scheme matches single-layer random neural networks \cite{Schmidt1992, Pao1994}.
Considering a $N \times L$ input dataset $\mathbf{X}$ and one output mode, the output function is
\begin{equation}
 \mathbf{Y}= \mathbf{H}\mathbf{\beta} = \sum_{j=1}^{M} H_j(\mathbf{X}) \beta_j ,
\end{equation}
where $\beta$ is the weight vector determined by training and $\mathbf{H(X)}=\mathbf{H}$ is the $N \times M$ hidden-layer matrix outcome.
$\mathbf{H(X)}=[\mathbf{g}(\mathbf{X}_1), ..., \mathbf{g}(\mathbf{X}_N) ]$, where
$\mathbf{g}(\mathbf{x})$  maps the input sample $\mathbf{x}$ from the $L$-dimensional input space to the $M$-dimensional feature space.
Under appropriate conditions, the machine is able to interpolate any continuous function (universal interpolator) and acts as a universal classifier \cite{Huang2012}. Given the target labels $\mathbf{T}$, training corresponds to solving the ridge regression problem: 
$\argminG_{\beta} ( \Vert \mathbf{H}\beta - \mathbf{T} \Vert^2 + c^{-1} \Vert \beta \Vert^2)$,
 where $c$ is a parameter controlling the trade-off 
between the training error and the regularization. The constrained optimization can be recast as a dual optimization problem \cite{Huang2012}.
A solution that is computationally affordable for large datasets is given by Eq. (1): 
$\beta= \left( \mathbf{H^{T}}\mathbf{H} +  c \mathbf{I} \right)^{-1} \mathbf{H^{T}} \mathbf{T}.$
In this case, matrix inversion involves the $M \times M$ matrix $\mathbf{H^{T}}\mathbf{H}$, which makes the method scalable 
and effective for large-scale applications.
The output function of the classifier is thus
\begin{equation}
\mathbf{Y}= \mathbf{H} \left( \mathbf{H^{T}}\mathbf{H} +  c \mathbf{I} \right)^{-1} \mathbf{H^{T}} \mathbf{T}.
\end{equation}
In the case of single-output node performing binary classification (as for the problem in Fig. 4(c)), 
the decision function is $f(\mathbf{X})=$ sign$(\mathbf{H(X)} \left[ \mathbf{H^{T}(X)}\mathbf{H(X)} +  c \mathbf{I} \right]^{-1} \mathbf{H^{T}(X)} \mathbf{T})$.
It generalizes for multi-class classifier with multiple output nodes as
$f(\mathbf{X})= \argmaxG_{k} \mathbf{Y}_k (\mathbf{X}) $ , where $\mathbf{Y}_k(\mathbf{X}) $ denotes the output $\mathbf{Y}$ of the $k$th nodes.

\subsection{Encoding methods} 

We consider phase encoding of the input data in all the presented results (phase-only light modulation).
The embedding matrix $\mathbf{W}$ is a fixed signal, independent of the specific input data, that
characterizes the phase encoder and modifies the PELM feature space. 
In the noise embedding method, $\mathbf{W}$ is chosen as a uniformly distributed random matrix with maximum amplitude $\rho$ (noise level) 
made by blocks of size $l$ (noise correlation length). 
In the Fourier embedding method we construct the embedding matrix $W=[W_1, ..., W_n]$ from $n$ frequencies, 
$W_k= \sum_{\omega=1}^{n} (a_\omega/n)\exp{(i\omega k/n)} $, 
with coefficients $a_\omega$ of equal amplitude and arbitrary phase.
Both the embedding signal and input data are encoded within the phase interval $[0,\pi]$.
In experiments, the preselected embedding matrix is discretized in gray levels and superimposed to each input data.

\section{APPENDIX B: EXPERIMENTAL SETUP AND DEVICE TRAINING}

A continuous-wave laser beam with wavelength $\lambda= 532$nm is expanded, polarized, and illuminates a reflective phase-only SLM (Hamamatsu X13138, $1280\times1024$ pixels, $12.5 \mu$m pixel pitch, $60$Hz maximum frame rate) that encodes input data within an embedding phase mask by pure phase modulation.
By grouping several SLM pixels, the modulator active area is divided into $L$ input nodes, with each node having $210$ phase levels equally distributed 
in the $0$-$2\pi$ interval. Phase-modulated light propagates in free space through a focusing plano-convex lens (f$=150$mm).
The optical field in the lens focal plane is collected by an imaging objective (NA$=0.4$) and detected by a CCD camera
with $8$-bit ($256$ gray levels) intensity sensitivity. To obtain a nonlinear saturated function of the transmitted field,
the camera exposure is manually increased to get overexposed images. 
$M$ output channels are preselected within the camera region of interest, where the signal in each channel is obtained by binning over few camera pixels
($10\times10$ for MNIST classification) to reduce detection noise.

Training is performed by loading one-by-one the $N$ input samples on the SLM, and keeping fixed the embedding matrix.
At each training step, values from the $M$ output channels are acquired (Fig. 3(c)) and stored on conventional computer controlling the setup.
The readout weights are obtained by applying Eq. (2) on the entire set of measurements.
They are readily used in the testing phase, where each of $N_t$ testing samples is send through the photonic machine and passively classified
by weighing the detected output.

\section{APPENDIX C: DATASETS FOR CLASSIFICATION AND REGRESSION}

Recognition of handwritten digits is tested on the MNIST dataset,
a standard benchmark for multiclass classification. The dataset, which includes $10$ classes, 
consists of $60000$ training sample ($N$) and $N_t= 10000$ digits for testing, each with size $L=28 \times 28$.
Although state-of-the-art convolutional neural networks reach accuracies exceeding $99.8$\% on MNIST (https://github.com/Matuzas77/MNIST-0.17), 
the task remains the basic test for any novel machine learning device. In fact, superior algorithms are application-specific and require 
massive data processing.

The \textit{mushroom} (https://archive.ics.uci.edu/ ml/ datasets/ Mushroom) is a binary class data sets with relatively large size and low dimension. 
It includes $8124$ samples with $L=22$ features in random order. The goal is to separate edible from poisonous mushrooms. 
A typical ELM accuracy is $88.9$ \% for a split ratio $N/N_t \approx 0.23$ \cite{Huang2012}. 

The \textit{abalone} dataset (https://archive.ics.uci.edu/ ml/ datasets/ Abalone) is one of the
mostly used benchmarks for machine learning, and concerns the identification of sea snails in terms of age and
physical parameters. Each training point has $L=8$ attributes, and the entire dataset has $8177$ samples.
Digital ELMs report testing errors around $0.07$ for $N/N_t \approx 2$. 
Errors are evaluated using the root mean square displacement RMSD$=\sqrt{ \sum_i^{N_t} ( Y_i  -  T_i )^2 /N_t }$. 

\vspace*{0.1cm}
{\bf Fundings.} We acknowledge funding from PRIN PELM 2017, QuantERA ERA-NET Co-fund (Grant No. 731473, project QUOMPLEX), H2020 PhoQus project (Grant No. 820392) and SAPIExcellence BE-FOR-ERC 2019.
\vspace*{0.1cm}

{\bf Acknowledgements.}
We thank MD Deen Islam and V.H. Santos for technical support in the laboratory.
\vspace*{0.1cm}

{\bf Disclosures.} 
The authors declare no conflicts of interest.


\vspace*{-0.1cm}


\begin{thebibliography}{70}

\bibitem{Haykin2008} S. Haykin, \emph{Neural networks and learning machines} (Pearson Prentice Hall, 2008).
\bibitem{Strubell2019} E. Strubell, A. Ganesh, A. McCallum, Energy and Policy Considerations for Deep Learning in NLP, { arXiv:1906.02243 } (2019).
\bibitem{Wetzstein2020} G. Wetzstein, A. Ozcan, S. Gigan, S. Fan, D. Englund, M. Soljačić, C. Denz, D.A.B. Miller and D. Psaltis, Inference in artificial intelligence with deep optics and photonics, {\it Nature\/} {\bf 588}, 39-47 (2020).
\bibitem{Farhat1985} N.H. Farhat, D. Psaltis, A. Prata, and E. Paek, Optical implementation of the Hopfield model, {\it Appl. Opt. \/} {\bf 24}, 1469 (1985).
\bibitem{Psaltis1988} D. Psaltis, D. Brady, and K. Wagner, Adaptive optical networks using photorefractive crystals, {\it Appl. Opt \/} {\bf 27}, 1752
(1988)
\bibitem{Denz1998} C. Denz, \emph{Optical Neural Networks} (Springer, Berlin, 1998).

\bibitem{Shen2017} Y. Shen, N.C. Harris, S. Skirlo, M. Prabhu, T. Baehr-Jones, M. Hochberg, X. Sun, S. Zhao, H. Larochelle, D. Englund, and M. Soljacic, 
Deep learning with coherent nanophotonic circuits, {\it Nat. Photon.} {\bf 11}, 441–446 (2017).
\bibitem{Lin2018} X. Lin, Y. Rivenson, N. T. Yardimci, M. Veli, Y. Luo, M. Jarrahi, and A. Ozcan, All-optical machine learning using diffractive deep neural networks, {\it Science\/} {\bf 361}, 1004–1008 (2018).
\bibitem{Engheta2019} N. Mohammadi Estakhri, B. Edwards, and N. Engheta, Inverse-designed metastructures that solve equations, {\it Science\/} {\bf 363}, 1333 (2019).
\bibitem{Bueno2018} J. Bueno, S. Maktoobi, L. Froehly, I. Fischer, M. Jacquot, L. Larger, and D. Brunner, Reinforcement learning in a large-scale photonic recurrent neural network, {\it Optica\/} {\bf 5}, 756–760 (2018).
\bibitem{Zuo2019} Y. Zuo, B. Li, Y. Zhao, Y. Jiang, Y. C. Chen, P. Chen, G. B. Jo, J. Liu, and S. Du, All-optical neural network with nonlinear activation functions, {\it Optica\/} {\bf 6}, 1132–1137 (2019).
\bibitem{Tait2017} A.N. Tait , T.F. de Lima, E. Zhou, A.X. Wu, M-A. Nahmias, B.J. Shastri, P.R. Prucnal, 
Neuromorphic photonic networks using silicon photonic weight banks, {\it Sci. Rep.\/} {\bf 7}, 7430 (2017).
\bibitem{Feldmann2019} J. Feldmann, N. Youngblood, C.D. Wright, H. Bhaskaran and W.H.P. Pernice, All-optical spiking neurosynaptic networks with self-learning capabilities, {\it Nature\/} {\bf 569}, 208-214 (2019).
\bibitem{Feldmann2021} J. Feldmann,  N. Youngblood, M. Karpov, H. Gehring, X. Li, M. Stappers, M. Le Gallo, X. Fu, A. Lukashchuk, A. S. Raja, J. Liu, C. D. Wright, A. Sebastian, T.J. Kippenberg, W.H.P. Pernice, and H. Bhaskaran, Parallel convolutional processing using an integrated photonic tensor core,
{\it Nature\/} {\bf 589}, 52-58 (2021).
\bibitem{Xu2021} X. Xu, M. Tan, B. Corcoran, J. Wu, A. Boes, T.G. Nguyen, S.T. Chu, B.E. Little, D.G. Hicks, R. Morandotti, A. Mitchell, and D.J. Moss, 11 TOPS photonic convolutional accelerator for optical neural networks, {\it Nature\/} {\bf 589}, 44-51 (2021).
\bibitem{Moss2020} X. Xu, M. Tan, B. Corcoran, J. Wu, T.G. Nguyen, A. Boes, S.T. Chu, B.E. Little, R. Morandotti, A. Mitchell, D.G. Hicks, and D.J. Moss, Photonic Perceptron Based on a Kerr Microcomb for High-Speed, Scalable, Optical Neural Networks, {\it Laser Photonics Rev.\/} 2000070 (2020).

\bibitem{Lvovsky2020} J. Spall, X.Guo, T.D. Barrett, and A.I. Lvovsky, Fully reconfigurable coherent optical vector-matrix multiplication, {\it Opt. Lett.
\/}{\bf 45}, 5752-5755 (2020).

\bibitem{Chang2018} J. Chang, V. Sitzmann, X. Dun, W. Heidrich, and G. Wetzstein, Hybrid optical-electronic convolutional neural networks with optimized diffractive optics for image classification, {\it Sci. Rep. \/} {\bf 8}, 12324 (2018).
\bibitem{Miscuglio2020} M. Miscuglio, Z. Hu, S. Li, J. George, R. Capanna, P.M. Bardet, P. Gupta, and V.J. Sorger, Massively Parallel Amplitude-Only Fourier Neural Network, {\it Optica\/} {\bf 7}, 1812 (2020).
\bibitem{Wu2021} C. Wu, H. Yu, S. Lee, R. Peng, I. Takeuchi, and M. Li, Programmable phase-change metasurfaces on waveguides for multimode photonic convolutional neural network, {\it Nat. Commun. \/}{\bf 12}, 96 (2021).

\bibitem{Hughes2019} T.W. Hughes, I. A. Williamson, M. Minkov, and S. Fan, Wave physics as an analog recurrent neural network, {\it Sci. Adv.\/} {\bf 5}, eaay6946 (2019).
\bibitem{Hamerly2019} R. Hamerly, L. Bernstein, A. Sludds, M. Soljačić, and D. Englund, Large-Scale Optical Neural Networks Based on Photoelectric Multiplication, Phys. Rev. X 9, 021032 (2019).
\bibitem{Hughes2018} T.W. Hughes, M.Minkov, Y.Shi, and S. Fan, Training of photonic neural networks through in situ backpropagation and gradient measurement, {\it Optica\/} {\bf 5}, 864-871~(2018).

\bibitem{Lukosevicius2009} M. Lukosevicius, and  H. Jaeger, Reservoir computing approaches to recurrent neural network training, 
{\it Computer Sci. Rev.\/} {\bf 3}, 127–149 (2009).
\bibitem{Gallicchio2017} C. Gallicchio, A. Micheli, L. Pedrelli, Deep reservoir computing: A critical experimental analysis, 
{\it Neurocomputing\/} {\bf 268}, 87-99 (2017).
\bibitem{Mattheakis2017} G. Neofotistos, M. Mattheakis, G.D. Barmparis, J. Hizanidis, G.P. Tsironis, and E. Kaxiras,  Machine learning with observers predicts complex spatiotemporal behavior, {\it Frontiers in Physics\/} {\bf 7}, 24 (2019).

\bibitem{Brunner2013} D. Brunner, M.C. Soriano, C. Mirasso, I. Fischer, Parallel photonic information processing at gigabyte per second data rates using transient states, {\it Nat. Comm.\/} {\bf 4}, 1364~(2013).
\bibitem{Vinkier2015} Q. Vinckier, F. Duport, A. Smerieri, K. Vandoorne, P. Bienstman, M. Haelterman, and S. Massar, High-Performance Photonic Reservoir Computer Based on a Coherently Driven Passive Cavity, {\it Optica\/} {\bf 2}, 438 (2015).
\bibitem{Larger2017} L. Larger, A. Baylón-Fuentes, R. Martinenghi, V.S. Udaltsov, Y.K. Chembo, and M. Jacquot, High-speed photonic reservoir computing using a time-delay based architecture: million words per second classification, {\it Phys. Rev. X\/} {\bf 7}, 011015 (2017).
\bibitem{Antonik2019} P. Antonik, N. Marsal, D. Brunner, and D. Rontani, Human action recognition with a large-scale brain-inspired photonic computer,
{\it Nat. Mach. Intell. \/}{\bf 1}, 530–537~(2019).

\bibitem{Rohm2020} A. Röhm, L. Jaurigue, and K. L\"udge, Reservoir Computing Using Laser Networks, IEEE J. Sel. Top. Quantum Electron. 26, 1 (2020).
\bibitem{Paudel2020} U. Paudel, M. Luengo-Kovac, J. Pilawa, T.J. Shaw, and G.C. Valley, Classification of time-domain waveforms using a speckle-based optical reservoir computer, {\it Opt. Express\/} {\bf 28}, 1225 (2020).
\bibitem{Rafayelyan2020} M. Rafayelyan, J. Dong, Y. Tan, F. Krzakala, and S. Gigan, Large-Scale Optical Reservoir Computing for Spatiotemporal Chaotic Systems Prediction, {\it Phys. Rev. X\/}{ \bf 10}, 041037 (2020).

\bibitem{Pierangeli2020} D. Pierangeli, V. Palmieri, G. Marcucci, C. Moriconi, G. Perini, M. De Spirito, M. Papi and C. Conti,
Living optical random neural network with three dimensional tumor spheroids for cancer morphodynamics, {\it Commun. Phys. \/}{\bf 3}, 1-10 (2020).

\bibitem{Huang2006} G.B. Huang, Q.Y. Zhu, and C.K. Siew, Extreme learning machine: Theory and applications, {\it  Neurocomputing \/}{ \bf 70}, 489 (2006).
\bibitem{Huang2012} G.B. Huang, H. Zhou, X. Ding, and R.Zhang, Extreme Learning Machine for Regression and Multiclass Classification, {\it IEEE 
Transactions on Systems, Man, and Cybernetics, Part B (Cybernetics)\/}{\bf 42}, 513-529 (2012).
\bibitem{Suykens1999} J.A.K. Suykens and J.Vandewalle, Least squares support vector machine classifiers, {\it Neural Process. Lett.\/} {\bf 9}, 293–300 (1999).
\bibitem{Schmidt1992} W.F. Schmidt, M.A. Kraaijveld, and R.P.Duin, Feed forward neural networks with random weights. In {\it International Conference on Pattern Recognition\/}, IEEE Computer Society Press, 1992. 
\bibitem{Pao1994} Y.H. Pao, G.H. Park, and D.J. Sobajic,  Learning and generalization characteristics of the random vector functional-link net, {\it Neurocomputing} {\bf 6}, 163 (1994).
\bibitem{An2007} S. An, W. Liu and S. Venkatesh, Face recognition using kernel ridge regression, in {\it CVPR 2007: Proceedings of the IEEE Computer Society Conference on Computer Vision and Pattern Recognition}, 2007. 

\bibitem{Saade2016} A. Saade, F. Caltagirone, I. Carron , L. Daudet, A. Dremeau, S. Gigan, F. Krzakala, Random projections through multiple optical scattering: approximating kernels at the speed of light, {\it IEEE International Conference on Acoustics, Speech and Signal Processing (ICASSP)\/} 6215–6219 (2016).
\bibitem{Uchida2020} S. Sunada, K. Kanno, and A. Uchida, Using multidimensional speckle dynamics for high-speed, large-scale, parallel photonic computing,
{\it Opt. Express \/}{\bf 28}, 30349 (2020).
\bibitem{Marcucci2020} G. Marcucci, D. Pierangeli, and C. Conti, Theory of neuromorphic computing by waves: machine learning by rogue waves, dispersive shocks, and solitons,  {\it Phys. Rev. Lett.\/} {\bf 125}, 093901 (2020).
\bibitem{Psaltis2020} U.Tegin, M.Yildirim, I.Oguz, C. Moser and D. Psaltis, Scalable Optical Learning Operator, arXiv:2012.12404~(2020).

\bibitem{Goodman2005} J.W. Goodman, \emph{Introduction to Fourier Optics} (Roberts and Company, 2005).

\bibitem{Huang2013} L.C.C. Kasun,H. Zhou, G. B. Huang, and C.M. Vong,  Representational learning with extreme learning machine for big data,
{\it IEEE intelligent systems\/}, {\bf 28}, 31 (2013).
\bibitem{Yan2019} T. Yan, J. Wu, T. Zhou, H. Xie, F. Xu, J. Fan, L. Fang, X. Lin, and Q. Dai, Fourier-space Diffractive Deep Neural Network, {\it Phys. Rev. Lett.\/} {\bf 123}, 023901 (2019).
\bibitem{Luo2019} Y. Luo, D. Mengu, N.T. Yardimci, Y. Rivenson, M. Veli, M. Jarrahi and A. Ozcan, Design of task-specific optical systems using broadband diffractive neural networks, {\it Light: Science \& Applications\/} {\bf 8}, 112 (2019).

\bibitem{Tzang2019} O. Tzang, E. Niv, S. Singh, S. Labouesse, G. Myatt, and R. Piestun,  Wavefront shaping in complex media with a 350 KHz modulator via a 1D-to-2D transform, {\it Nat. Photonics} {\bf 13}, 788 (2019).
\bibitem{Braverman2020} B. Braverman, A. Skerjanc, N. Sullivan, and R.W. Boyd, Fast generation and detection of spatial modes of light 
using an acousto-optic modulator, {\it Opt. Express\/} {\bf 28}, 29112 (2020).

\bibitem{Piccinotti2020} D. Piccinotti, K.F. MacDonald, S. Gregory, I. Youngs and N.I. Zheludev, Artificial Intelligence for Photonics and Photonic Materials, {\it Rep. Prog. Phys. \/} (2020). DOI:10.1088/1361-6633/abb4c7
\bibitem{Genty2020} G. Genty, L. Salmela, J.M. Dudley, D. Brunner, A. Kokhanovskiy, S. Kobtsev and S.K. Turitsyn,
 Machine learning and applications in ultrafast Photonics, {\it Nat. Photonics\/} (2020). DOI:/10.1038/s41566-020-00716-4
\bibitem{Kutz2017} S.H. Rudy, S.L. Brunton, J.L. Proctor, and J.N. Kutz, Data-driven discovery of partial differential equations, {\it Sci. Adv.\/} {\bf 3}, e1602614 (2017).

\bibitem{McMahon2016} P.L. McMahon, A. Marandi, Y. Haribara, R. Hamerly, C. Langrock, S. Tamate, T. Inagaki, H. Takesue, S. Utsunomiya, K. Aihara,
R.L. Byer, M.M. Fejer, H. Mabuchi, Y. Yamamoto, A fully-programmable 100-spin coherent Ising machine with all-to-all connections, {\it Science\/} {\bf 354}, 614 (2016).
\bibitem{Inagaki2016} T. Inagaki, Y. Haribara, K. Igarashi, T. Sonobe, S. Tamate, T. Honjo, A. Marandi, P. L. McMahon, T. Umeki, K. Enbutsu,
O. Tadanaga, H. Takenouchi, K. Aihara, K.-i. Kawarabayashi, K. Inoue, S. Utsunomiya, H. Takesue, A coherent Ising machine for 2000-node optimization problems, {\it Science\/} {\bf 354}, 603 (2016).
\bibitem{Vandersande2019} F.B\"ohm, G. Verschaffelt, and G. Van der Sande, A poor man’s coherent Ising machine based on opto-electronic feedback systems for solving optimization problems, {\it Nat. Commun.\/} {\bf 10}, 3538 (2019).
\bibitem{Pierangeli2019} D. Pierangeli, G. Marcucci, and C. Conti, Large-Scale Photonic Ising Machine by Spatial Light Modulation, {\it Phys. Rev. Lett.\/}
{\bf 122}, 213902 (2019).
\bibitem{Pierangeli2020_2} D. Pierangeli, G. Marcucci, and C. Conti, Adiabatic evolution on a spatial-photonic Ising machine, {\it Optica \/}{\bf 7}, 
1535~(2020).
\bibitem{Prabhu2020} M. Prabhu, C. Roques-Carmes, Y. Shen, N. Harris, L. Jing, J. Carolan, R. Hamerly, T. Baehr-Jones, M. Hochberg, V. Čeperić,
J.D. Joannopoulos, D.R. Englund, and M. Soljačić, Accelerating recurrent Ising machines in photonic integrated circuits, {\it Optica\/} {\bf 7}, 551 (2020). 
\bibitem{Kalinin2020} K.P. Kalinin, A. Amo, J. Bloch, and N.G. Berloff, Polaritonic XY-Ising Machine, {\it Nanophotonics\/}{\bf 9}, 4127 (2020).
\bibitem{Gaeta2020} Y.Okawachi, M.Yu, J.K.Jang, X.Ji, Y.Zhao, B.Y.Kim, M.Lipson and A.L. Gaeta, Demonstration of chip-based coupled degenerate
optical parametric oscillators for realizing a nanophotonic spin-glass, {\it Nat. Commun.\/} {\bf 11}, 4119~(2020).
\bibitem{DiFalco2019} A. Di Falco, V. Mazzone, A. Cruz and A. Fratalocchi, Perfect secrecy cryptography via mixing of chaotic waves in irreversible time-varying silicon chips, {\it Nat. Commun.\/} {\bf 10}, 5827 (2019).


\end{thebibliography}
\end{document}